\documentclass[12pt]{article}

\usepackage{scicite}

\usepackage{times}

\usepackage{color}
\usepackage{graphicx}

\usepackage[all,cmtip]{xy}

\newenvironment{sciabstract}{%
\begin{quote} \bf}
{\end{quote}}

\newcommand{\pnA}{P$^{\mathrm{N}}$}
\newcommand{\poxA}{P$^{\mathrm{OX}}$}

\newcommand{\ppA}{P$^{\mathrm{1+}}$}
\newcommand{\pn}{P$^{\mathrm{N}}$ }
\newcommand{\pox}{P$^{\mathrm{OX}}$ }

\newcommand{\pp}{P$^{\mathrm{1+}}$ }

\newcommand{\rmFe}{\mathrm{Fe}}

\newcommand{\vcr}{\vec{r}}

\newcommand{\rSeff}{S_{\mathrm{eff}}}

\newcommand{\ceFeFour}{Fe$_4$S$_4$}
\newcommand{\ceFeEight}{Fe$_8$S$_7$}

\newcounter{lastnote}

\title{Electronic landscape of the P-cluster of nitrogenase
as revealed through many-electron quantum wavefunctions}

\author
{Zhendong Li$^{1\ast}$, Sheng Guo$^1$, Qiming Sun$^1$, Garnet Kin-Lic Chan$^{1\ast}$\\
\normalsize{Division of Chemistry and Chemical Engineering, California Institute of Technology,}\\
\normalsize{Pasadena, CA 91125, USA}\\
\normalsize{$^\ast$To whom correspondence should be addressed;}\\
\normalsize{E-mail: zhendongli2008@gmail.com, gkc1000@gmail.com}}

\date{}

\begin{document}


\baselineskip24pt

\maketitle

\begin{sciabstract}
  The electronic structure of the nitrogenase metal cofactors is central to nitrogen fixation.
  However, the P-cluster and iron molybdenum cofactor, each containing eight irons, have
  resisted detailed characterization of their electronic properties.
  Through exhaustive many-electron wavefunction simulations enabled by new
  theoretical methods,  we report on the low-energy electronic states of the P-cluster
  in three oxidation states. The energy scales of orbital and
  spin excitations overlap, yielding a dense spectrum with features we  trace to the underlying
  atomic states and recouplings.  The clusters exist in superpositions
  of spin configurations with non-classical spin correlations,
  complicating interpretation of magnetic spectroscopies,
  while the charges are mostly localized from reorganization of the cluster and its surroundings. Upon
  oxidation, the opening of the P-cluster significantly increases the density of states,
  which is intriguing given its proposed role in electron transfer.
  These results demonstrate that many-electron simulations
  stand to provide new insights into the electronic structure of the nitrogenase cofactors.
  \end{sciabstract}


The Fe-S clusters of nitrogenase, namely, the [\ceFeFour] Fe-cluster of the Fe-protein,
and the [\ceFeEight] P-cluster and [MoFe$_7$S$_9$C] FeMo-cofactor of the MoFe protein, are the active sites for electron transfer and
reduction in biological nitrogen fixation\cite{beinert1997iron,howard1996structural,rees2003interface,hoffman2014mechanism}.
The P-cluster and FeMo-cofactor, in particular, stand as peaks
in the electronic complexity of enzymatic cofactors, with eight spin-coupled, open-shell, transition metal ions.
Resolving their atomic and electronic structure has stood as a major challenge for
experimental and theoretical spectroscopies.
In the last decades, careful application of experimental techniques, including
X-ray crystallography and spectroscopy, M\"ossbauer spectroscopy, and electron paramagnetic resonance (EPR), amongst
others, has led to precise atomic structures, first for the Fe-cluster, followed by the P-cluster,
and most recently, the FeMo-cofactor\cite{spatzal2011evidence,lancaster2011x}.
Like others, we have been working towards a complementary goal of
determining the electronic structure
of the three clusters at a detailed, many-electron, level. A few years ago, we reported an \emph{ab initio} picture of the
electronic states of the (4Fe-4S) Fe cluster
that revealed a rich low-energy landscape\cite{sharma_low-energy_2014}, with a density and variety of states far greater than previously thought.
In this article, we report on the electronic landscape of the P-cluster.

The P-cluster is thought to mediate electron transfer from the Fe-cluster to the FeMo-cofactor where nitrogen reduction occurs.
The relevant charge states are considered to be the resting state \pnA, the one-electron oxidized state \ppA, and the two-electron oxidized state \poxA
~\cite{chan1999spectroscopic,danyal2011electron}, although the recently proposed deficit spending mechanism\cite{danyal2011electron,seefeldt2018energy} postulates
that only the {\pnA} and {\ppA} states take part in electron transfer.
X-ray crystallography reveals a structure consisting of an [\ceFeEight] core and two cubanes sharing a sulfur bridge. Under oxidation
the cluster undergoes a large structural opening, driven by coordination
first of Ser-$\beta$188 to Fe4 in \ppA, followed by Cys-$\alpha$88 to Fe2 in \poxA~\cite{peters1997redox,keable2018structural}; the
role of this structural opening in the mechanism of electron transfer is poorly understood. Fig. 1a highlights the
redox-dependent structural rearrangements across \pnA, \ppA, and \pox.

While the atomic structure of the P-cluster is resolved, information on
the electronic structure is  fragmentary (see SM for a brief introduction).
However, relating structure to chemical function requires to know the
basic oxidation and spin states, and interpreting spectra, particularly in magnetic spectroscopies such as EPR,
requires models of the low-energy, spin-coupled, many-electron quantum states, and their spin and charge distributions.
Here, we describe the first \emph{ab initio} calculations of the quantum states of the P-cluster at
the many-electron level. By some measures, the electronic complexity of this system is greater than that of
any other molecular problem studied with \emph{ab initio} many-electron simulation.
In particular, the low-lying electronic states of the P-cluster are more numerous, and the spectrum
is denser, than what we previously encountered in the already challenging Fe-cluster.
As a result, the energy resolution we can achieve ($<$3 kcal/mol in energy differences)
does not fully resolve the precise ordering of all states.
However, analogous to how an X-ray crystal structure at close to atomic-scale resolution displays the larger scale structural features of a protein, our current calculations reveal the larger scale features of the electronic landscape.
In particular, we wish to understand how spin and charge is distributed in the P-cluster;
how the individual moments combine to form the global spin state; what types of excitations appear at low energies;
and the implications for spectroscopy and the function of the P-cluster in electron transfer.
We will  address these questions using new theoretical techniques and exhaustive simulations.


\section*{Theoretical strategy}

We first recall why \emph{ab initio} many electron simulation is necessary even to achieve a qualitatively correct description.
One-electron computational models, such as provided by broken-symmetry density functional theory (BS-DFT)\cite{noodleman1985models,noodleman1986ligand,yamaguchi1989antiferromagnetic,shoji2006theory}, are commonly
applied with success to transition metal complexes.
However, they encounter limitations when there are multiple open-shells with mixed valence character, as found in the P-cluster.
This is because BS-DFT requires a fixed spin configuration at each Fe center, and thus does not treat
the spin-coupling between centers that produces the global electronic spin-state
(with a well-defined spin quantum number) as a superposition of many different
metal spin configurations. Instead, to handle this, BS-DFT
is usually augmented by a model analysis,
where a parametrized model spin-Hamiltonian is diagonalized to obtain the many-spin wavefunction (a simplification of the true many-electron
state). However, as we determined in the [\ceFeFour] cluster~\cite{sharma_low-energy_2014},
such model Hamiltonians do not properly account for the multiple $d$ orbitals
relevant to mixed valence metals, and thus miss many of the low-lying excited states, underestimating the spectral density in the [\ceFeFour] cluster by an order of magnitude or more.
A complete picture can thus only be  obtained when superpositions of both spin and orbital degrees of freedom
are fully considered; this is what is provided by an \emph{ab initio} many-electron simulation.

Solving the many-electron Schr\"odinger equation exactly for the full P-cluster is not currently possible.
To obtain a tractable simulation, we instead solve the Schr\"odinger equation
only within the space of the most important orbitals, a complete active space (CAS), and represent
the wavefunction as a matrix product state (MPS)\cite{schollwock_density-matrix_2011}, the class of wavefunctions generated by the
density matrix renormalization group (DMRG)\cite{white_ab_1999,chan_density_2011}.
In iron-sulfur clusters, the natural active space comprises the Fe $3d$ and S $3p$ (necessary
for double exchange~\cite{anderson1955considerations}) valence orbitals.
However, even within the active space, and only counting configurations derived
from ferric Fe $3d$ configurations, the number of relevant electron configurations is enormous,
with as many $\sim 10^{22}$ relevant configurations for the [\ceFeEight] core.
To tackle this, the MPS provides a compression of the wavefunction, controlled by single parameter $D$, the bond dimension. Simulating with an MPS with different finite $D$, followed by extrapolating  $D\to \infty$, then yields an estimate of the exact solution
in the active space.

We previously used such CAS DMRG simulations to correctly describe the electronic states of
the Fe-cluster\cite{sharma_low-energy_2014}.
However, there are additional complications in the P-cluster that require special theoretical treatment,
arising from the fact that the low-energy spectrum contains
a large variety of qualitatively very different electronic states. As a result, the
non-linear optimization of the wavefunction in the DMRG algorithm is easily trapped in local minima, preventing access to the
full spectrum. To address this, we have designed a new theoretical procedure that combines the simple enumerability of BS-DFT solutions
with the representational power of MPS. We first enumerate
all low-lying BS-DFT states, each of which can be thought of as a simple, $D=1$, MPS. From these
electronic basins, we then apply spin-projection and increase the bond-dimension of the MPS to obtain a more flexible wavefunction that can properly capture the many-electron eigenstate. The procedure is outlined schematically in Fig. 1c
and described in detail in the SM.

Using this procedure, we carried out calculations on structural models of the \pnA, \ppA, \pox clusters
derived from the cofactor in {\it Azotobacter vinelandii} and its surrounding residues,
as well as a synthetic model cluster~\cite{ohki2003synthesis} (Fig. 1a). The \emph{ab initio} CAS DMRG calculations treated up to 120 electrons in 77 orbitals CAS(120e,77o).
Further details are in the methods section as well as in the SM.

It is important to estimate the sources of theoretical uncertainty.
There are multiple sources including the geometry and environment,
basis sets, correlation treatments, and the convergence of the DMRG solutions.
From a detailed analysis of these factors in smaller chemical analogs (see SM),
we estimate the uncertainty in the relative energies of the low-lying states of
interest (within ca. $10$ kcal/mol of the ground state) at a fixed geometry to be less than 3 kcal/mol
in most cases, stemming from the uncertainty in the basis set and correlation treatments (ca. 2 kcal/mol) as well as
in the DMRG convergence ($<$0.8 kcal/mol, except for the uncertainty in the energy
difference between the lowest \pox $S=3$ and $S=4$ states, which is about
2.8 kcal/mol).
This level of accuracy means that the low energy states we observe will remain
low energy states in more exact calculations, although it does not resolve all state orderings due
to the high density of the spectrum. Consequently, our objective is to address qualitative questions regarding the structure of the low energy
spectrum.
The electronic landscape we uncover serves as a zeroth order approximation for more elaborate calculations in the future.

\section*{Results}

\subsection*{Families of low-lying states in the P-cluster}

Before proceeding to the computational results, we briefly present
a framework to understand the electronic landscape (Fig. 2).
The low energy electronic states of the P-cluster (i.e. states within a few kcal/mol of the ground-state)  arise from
different ways to assemble the global state from individual Fe configurations.
The relevant Fe oxidation states lie between Fe(II), Fe(III), with corresponding
spins between $S=2$ and $S=5/2$. The orbital configurations arise
from an approximately tetrahedral, weak ligand field from the sulfurs
around each metal, with distortions that further split the $t_2$ and $e_g$ orbitals\cite{coucouvanis1981tetrahedral}.

Since the P-cluster consists of conjoined cubanes, the [\ceFeFour] cubane
is a natural intermediate unit to understand the electronic structure. Each cubane itself has
a large number of states at low energy, arising from
(1) local ligand-field excitations of Fe(II)/Fe(III) ions,
(2) double exchange hopping between all the Fe $3d$ orbitals,
and (3) Heisenberg-like ladders of spin-coupling in Fe-dimers (and dimers of such dimers).
Double exchange couples the spin state and charge delocalization in the cubane~\cite{noodleman1991exchange}.

The states of the two cubanes are further electronically coupled across the bridging sulfur
to produce the global P-cluster state. Note that the coupling not only gives rise to families of states (such
as Heisenberg ladders generated from the two cubane total spins) but also allows for different types of states on the same cubane to mix.
This latter case gives a true many-electron ``multi-configurational'' superposition, as one cannot
assign a single spin configuration (i.e. configuration state function) to the left or right cubane.
Instead, the global state needs to be described by a superposition of couplings
of various different left and right cubane states, and we can only refer to
the average cubane spins ($S_{\mathrm{eff}}$ or $S_L$, $S_R$), computed as expectation values, for the cubanes.

From the above scheme, we can group the low-energy states into 3 families (Fig. 2). The
first are (A) spin isomers, which have the same total spin as the ground-state, but differ by local reorientations of the spins and their couplings;
these are conveniently discussed in terms of changes in the underlying cubane states; (B) orbital
excitations, where electron configurations of the ions change (e.g. in ligand field excitations); and (C) global spin excitations, where the cubanes
recouple into a different spin state from the ground-state. Excitations in class A and C
are true many-electron excitations and cannot be fully described within BS-DFT, while excitations in class B are missed in typical model Hamiltonians, which do not contain all $3d$ orbital degrees of freedom. The energy scales associated with A are the exchange and double-exchange
energies within a cubane; that associated with B is the ligand field splittings; and that associated with C is the exchange and double
exchange between the cubanes. As we shall see, these energy scales all overlap in the low-energy states of the P-cluster.

\subsection*{The \pn cluster and its synthetic model}

\noindent \textit{Ground states}. We start with the ground states of \pn and a synthetic model cluster
with a similar core geometry (Fig. 3).
Consistent with M\"ossbauer and EPR spectra, we find the \pn ground state to be a singlet ($S=0$) with
all ferrous irons. The cubanes assume $S=4$ states ($\rSeff\approx 3.7$)
corresponding to the $S=-2+(2+2+2)=4$ (3:1) coupling scheme between the terminal and other three Fe's.
We note that this cubane electronic motif is also found in the  [\ceFeFour]$^0$ cluster in the super-reduced iron protein\cite{watt1994formation,angove1997mossbauer}, and
cubanes in similar electronic states have been used to assemble the P-cluster in a non-enzymatic synthesis\cite{rupnik2014nonenzymatic}.

In contrast to the natural \pn cluster, the synthetic model also has an $S=0$ ground-state, but with
Fe oxidation states close to 6Fe(II)2Fe(III). However, the cubane electronic states are related to those in the \pn cluster,
with the charges localized on
the terminal Fe's, and
a similar 3:1 coupling pattern yielding  $S=7/2$ cubanes. Note that the driving force for charge localization does not come solely from the terminal ligands,
but also has a small spin-coupling contribution: a delocalized
cubane charge is typically associated with an $S=1/2$ cubane~\cite{noodleman1991exchange}, but the $S=7/2$ cubane (with localized charge) maximizes
the number of antiferromagnetic (AFM) Fe pairs. In particular, in the
$S=7/2$ cubane, there are 9 (intercubane) + 6 (intracubane) = 15 AFM pairs,
while for the $S=1/2$ cubane, there are 5 (intercubane) + 8 (intracubane) = 13 AFM pairs.


\noindent \textit{Excited states}. All three classes of excited states are found in the low-energy ($<8$ kcal/mol) spectrum (Figs. 3a, 3b).
Global recoupling of the $S=4$ (\pnA) (model: $S=7/2$) cubanes into an overall triplet ($S=1$) state occurs at 0.6 kcal/mol (\pnA) (model: 1.1 kcal/mol) above the ground-state. This leads to an inter-cubane Heisenberg ladder of states (class C excitations) with an effective $J \approx 200\mathrm{cm}^{-1}$.
Intracubane spin (class A) excitations that modify the 3:1 coupling on one of the cubanes occur at higher energies (e.g. starting at $\sim 6$ kcal/mol)
reflecting the larger intracubane effective $J$.
Finally, some ligand-field excitations from partially split $e$ and $t_{2}$ orbitals appear at very low energy and on a similar
scale to the spin excitations, with the lowest $d\rightarrow d$ transition in Fe8(II) appearing at about 2 kcal/mol (ca. 700cm$^{-1}$).
Both localized ligand field excitations, as well as delocalized linear combinations of $d\to d$ transitions similar
to those in bulk FeS, can be found.

\subsection*{The oxidized P-clusters: \pp and \pox}

In the \pp and \pox clusters, the opening of the core weakens the coupling between the cubanes.
Together with the multiple Fe oxidation states and overall non-singlet ground-state,
this gives rise to a greater degeneracy
of states at low energy and a more complicated spectrum. The sets of \pp and \pox  states
can be related both to each other and to those of \pnA, and these
relationships are illustrated in Fig. 5.


\noindent \textit{$S=1/2$ and $S=5/2$ ground-states of {\ppA}}.  EPR suggests that both $S=1/2$ and $S=5/2$ are potential ground state spins for \ppA~\cite{tittsworth1993detection}. We find the $S=5/2$ ground state to be
slightly lower in energy than the $S=1/2$ state (-2.7 kcal/mol) but within
the range of theoretical uncertainty.
In both the $S=1/2$ and $S=5/2$ ground-states, the charge is localized, with oxidation mainly on Fe4,
which coordinates to the oxygen in Ser-188 in conjunction with the structural opening.
Additionally, the irons in the right cubane are overall  more reduced than the irons in the left cubane (shading in Figs. 4a, 4b)
likely due to the compressed cubane structure and closer proximity of the ligands, furthering the charge imbalance between the cubanes.
The spin structure of the \pp cluster ground-state is complicated (Figs. 4a, 4b): in the
$S=1/2$ (A1) ground-state, the left cubane
consists of a ferromagnetically coupled (dimer of $S\approx 4$ dimers)  $S_L=4.1$ state, with
the right cubane in a similar $S_R\approx 4$, 3:1 coupled state as seen in the \pn cluster.
In the $S=5/2$ (A1) ground-state, the left cubane
consists of the same dimer of dimers, but now coupled antiferromagnetically into a low spin $S_L=0.9$ state,
with the right cubane in a superposition of cubane states with $S_R=2.3$, see Fig. 5.

\noindent \textit{$S=3$ and $S=4$ ground-states of \poxA}. 
Experimentally, the ground-state spin of the \pox cluster is not known with certainty\cite{surerus1992mossbauer}, and both $S=3$ and $S=4$ spin states can be identified in related P-clusters\cite{owens2016tyrosine}, such as in {\it Gluconacetobacter diazotrophicus}.
For $S=4$ we find that the ground-state is close to degenerate (states A1, A2),
while the $S=3$ (A1) ground state lies above the $S=4$ ground-state by 1.4 kcal/mol, but well
within the theoretical uncertainty.
In both spin states, Fe2 and Fe4 are the principal sites of oxidation, with
largely localized charges correlated with
ligating to hard ligands (O and N) and the structural opening, with some residual ferric character
on the other Fe's in the left cubane (shading in Figs. 4c, 4d). The opening of the left cubane
resembles that seen in isolated cubanes under oxidation, such as
the conformational change of the [Fe$_4$S$_3$] cluster upon oxidation in oxygen-tolerant membrane-bound [NiFe] hydrogenases\cite{shomura2011structural,fritsch2011crystal,volbeda2012x,tabrizi2015mossbauer}.
Unlike in the \pn cluster, there is no simple classical picture of the spin-coupling in \poxA; each state
is spin-canted\cite{noodleman2006structure} and involves a linear combination of multiple spin coupling schemes.


\noindent \textit{Excited states of \pp and \poxA}. As shown in Figs. 4c, 4d, similarly to in \pnA, we find localized and delocalized low-energy $d\to d$ ligand field transitions (class B excitations). The main difference between the \pp and \pox low energy landscape from that of \pn arises from the spin isomers (class A and C excitations).
Whereas in the \pn cluster we find that there is some separation in energy scales between  inter-(small) and intra-(large) cluster spin reorganizations,
in the \pp and \pox clusters both energy scales are comparably small, leading to a higher density of states. Thus while our procedure
identifies only a spin-coupling pattern of the \pn cluster ($S=0$) below 5 kcal/mol, in the \pp cluster we find 2 and in the \pox cluster
we find 4. In particular, the opening of the left cubane and additional charges in \pp and \pox
introduces new left cubane coupling patterns (2:2 or dimer of dimers) at comparable energies to the 3:1 spin-coupled state in \pn. The
dimer of dimers spin coupling motif leads to an approximate spin ladder in the left cubane ($S_L\approx 0.8,1.6,2.4,3.3,4.1$) that can be seen in the different low energy P-cluster states (see left panel of Fig. 5).
The fractional $S_L$ is due to charge delocalization (see Fe in dark red in the inset of Fig. 4).
The right cubane states can be found in various combinations of 2:2 and 3:1 spin-coupling patterns. Fig. 5 shows how these different excited states arise from the underlying cubane spin states in the P-clusters.


\section*{Discussion}


\noindent \textit{The cluster states can display non-classical spin correlations.}
From a theoretical perspective, it is useful to highlight the electronic features of the states we observe
beyond that described in standard treatments.
For example, even though the BS-DFT spin densities would be qualitatively wrong in all states
here (e.g. giving non-vanishing spin density in the $S=0$ states) the \textit{relative} orientation of the iron spins
(a so-called ``classical'' spin correlation pattern)
could in principle still resemble that in the many-electron state.
In some states, such as the $S=0$ ground state (A1) of \pn
and $S=1/2$ A2 state of \ppA, this is indeed the case. However,
a succinct example of a state with non-classical spin correlations is the $S=3$ ground state of \pox (see Fig. 4c) which has
6 spin-up and 2 spin-down Fe's. Within BS-DFT, such a spin distribution can {\it only} arise for an $S>3$ state, leading to an inconsistency.
Similarly, compared to a traditional model Hamiltonian analysis, the cubane effective spins $S_L$ and $S_R$ away from integer or half-integer cannot
be captured by a simple spin-coupling scheme. These kinds of non-classical spin correlations
clearly affect the interpretation of magnetic spectroscopies (see below), which have previously relied on BS-DFT or
a single spin-coupling scheme.
It may also have implications for the electronic structure of the FeMo-cofactor, where analysis of
 the $S=3/2$ ground state spin distribution has so far exclusively relied on BS-DFT~\cite{lovell2001femo,dance2010electronic,siegbahn2016model,bjornsson2017revisiting,cao2018influence}.

\noindent  {\textit{The oxidized clusters contain localized charges}}. Despite the multiple metal centers, in \pp and \pox the
charges are strongly localized and distributed asymmetrically between the cubanes. This is consistent with strong coupling to the ligands,
a polaronic effect, and is reflected   in the large geometric rearrangements of the cluster. We note that localized charge distributions have also
  been observed recently in FeMo-cofactor~\cite{spatzal2016nitrogenase}. This localization may be related to controllability of the density of states (see below).

\noindent \textit{Energy scales of spin and electronic excitations are comparable}. While
ligand field splitting and exchange effects are usually associated with
 different energy scales, both spin and orbital excitations
 are found at low energies in the P-cluster, especially in \pp and \poxA.
 Spectroscopically, this means that apparently local magnetic responses need not arise
 from spin localisation. For example, recent MCD measurements found similar spectral features in \pp and
  the [\ceFeFour]$^+$ cubane\cite{rupnik2012p+}, which was interpreted as meaning that the spin density in \pp is localized to
  a single cubane. However, as expected for a non-singlet ground-state, the spin density in \pp is
  actually distributed across {\it both} cubanes. Instead, the similarity between \pp and the [\ceFeFour]$^+$ cubane MCD spectrum
  is more naturally explained in terms of the localized nature of the cubane one-electron excitations (class B excitations), as evidenced
  by the ligand field excitations, which are also probed by the MCD response.

\noindent {\textit{Hyperfine coupling analysis.}  Using spin distributions consistent with a many-electron wavefunction representation of the states,
  we can attempt to reinterpret hyperfine coupling parameters using
  a model previously developed in~\cite{mouesca1994analysis,mouesca1995spin}.
  In \poxA, hyperfine coupling parameters have been reported for some time, but
  have previously been analyzed only in terms of spin distributions consistent with
  a single configuration state function description\cite{mouesca1994analysis}.
  Note that the parameters are themselves obtained by fitting and are thus uncertain, but we can compare
  at least the relative signs and magnitudes of the internal magnetic field $H_{\mathrm{int}}$ to the local spin $S_{\rmFe}^z$ for qualitative insight. As shown in Table \ref{tab:pOXcomparison}, the hyperfine parameters have a 6:2 sign pattern similar to that of the $S=3$ ground state spin distribution (that arises from the multi-configurational superposition) whereas the lowest $S=4$ state
  has a 5:3 sign pattern.
Unfortunately, the 5:3 pattern was not considered in earlier experimental fits\cite{huynh1980nitrogenase}, and thus cannot be compared on an equivalent footing.
Out of the two 6:2 patterns,
the small  $S_{\rmFe}^z=0.47$ for Fe2 (ligated with the
amide-N) correlates with the small hyperfine parameter no. 4, 
while the increased charge on Fe2, which thus has more ferric character compared to the other sites,
is consistent with the isomer shift of $\delta=0.25$~mm/s for the same iron (no. 4).
The \pox $S=3$  ground state thus correlates best with the experimentally reported hyperfine parameters.

\noindent {\textit{The density of states increases significantly upon oxidation}}. Near the ground-state, the density of states in \pp and \pox is much} higher than that of \pnA, the physical origin of which can be traced to the opening of the cluster.
This is intriguing in the context of the role of P-cluster in the electron transfer process.
In the proposed deficit spending mechanism\cite{danyal2011electron,seefeldt2018energy}, the first step is a slow electron transfer from the \pn cluster to the FeMo-cofactor, followed by a fast refilling of the oxidized P cluster from the Fe cluster. The higher density of states in \pp is thus consistent with the fast refilling.

\noindent {\textit{Controllable states hypothesis}}.
The P-cluster has a large density of states,  a factor that affects electron transfer.
There are in principle two ways to obtain such a large number of states.
The first is to create a (partially filled) metallic band from strong orbital overlap. The second is to have a
large number of weakly coupled, localized atomic orbitals at different energies, arising from different local environments.
The latter is much closer to the situation in the P-cluster (and Fe-S clusters in general).
We suggest this second scenario, which gives rise to complex spin couplings from the many-electron
recoupling of local states, could be favoured in the nitrogenase clusters because it provides for greater controllability.
The local origin of the states means that local changes in geometry (or environment, such as ligation) can be used to modify
the density of states, as we see in the coupling of the low-energy landscape to the opening of the
[\ceFeEight] core structure as one progresses from \pn to \pp to \pox (Fig. 5).


\begin{table}\centering
\caption{Comparison of the experimental hyperfine parameters
(internal magnetic fields $H_{\mathrm{int}}$ in kG,
quadrupole splittings $\Delta E_Q$ in mm/s,
isomer shifts $\delta$ in mm/s) from Ref. \cite{huynh1980nitrogenase} with the computed local charges $N_{\rmFe}$ and spin projections $S_{\rmFe}^z$
for the lowest $S=3$ and $S=4$ states of the \pox cluster of \emph{Azotobacter vinelandii}. The values of $N_{\rmFe}$ for ground states of \pn
and the synthetic model are listed as references.}\label{tab:pOXcomparison}
\begin{tabular}{ccccccccccccc}
\hline\hline
\multicolumn{4}{c}{Expt.\cite{huynh1980nitrogenase}} && \multicolumn{2}{c}{\pox ($S=3$)} && \multicolumn{2}{c}{\pox ($S=4$)} && \multicolumn{2}{c}{$N_{\rmFe}$} \\ \cline{1-4}\cline{6-7}\cline{9-10}\cline{12-13}
no. & $H_{\mathrm{int}}$ & $\Delta E_Q$ & $\delta$ & atom$^a$
& $N_{\rmFe}$ & $S_{\rmFe}^z$ && $N_{\rmFe}$ & $S_{\rmFe}^z$ &&
 \pnA & Syn \\
\hline
1	&	-287	& -1.40 & 	0.56 	&		Fe1	&	6.24 	&	0.52 	&	&	6.24 	&	-0.52 	&	&	6.33 	&	6.15 \\
2	&	-237	& +1.53 &	0.64 	&		Fe2	&	6.15 	&	0.47 	&	&	6.15 	&	0.44 	&	&	6.46 	&	6.32 \\
3	&	-245	& +0.57 &	0.40 	&		Fe3	&	6.22 	&	0.55 	&	&	6.23 	&	0.43 	&	&	6.38 	&	6.30 \\
4	&	-151	& +0.60 &	0.25 	&		Fe4	&	6.07 	&	1.04 	&	&	6.08 	&	-0.58 	&	&	6.35 	&	6.44 \\
5	&	-259	& +1.26 &	0.48 	&		Fe5	&	6.34 	&	-1.24 	&	&	6.32 	&	1.64 	&	&	6.51 	&	6.30 \\
6	&	-237	& -0.72 &	0.49 	&		Fe6	&	6.41 	&	1.42 	&	&	6.40 	&	1.66 	&	&	6.48 	&	6.32 \\
7	&	201	& +3.20 &	0.65 	&		Fe7	&	6.33 	&	1.55 	&	&	6.28 	&	1.74 	&	&	6.33 	&	6.45 \\
8	&	223	& +2.30 &	0.68 	&		Fe8	&	6.18 	&	-1.40 	&	&	6.26 	&	-1.33 	&	&	6.32 	&	6.15 \\
\hline\hline
\multicolumn{13}{l}{$^a$ The index does not correspond to the no. in the first column.}
\end{tabular}
\end{table}

\section*{Conclusions}

In summary, we have shown that we can now access the electronic structure of the P-cluster of nitrogenase
at the level of many-electron wavefunction simulations that can qualitatively capture the richness
of the low energy landscape. We report on a plethora of low-energy states across the  \pnA, \ppA, and \pox clusters and
their composition in terms of the local atomic configurations,  spins, and spin-couplings.
The many-electron nature of the states manifests in unusual spin distributions that are not contained
in simpler broken-symmetry analyses, and energy scales of spin and charge excitations that overlap. Together these imply a non-trivial
interpretation of magnetic spectroscopies. The opening of the cluster leads to an increase in the low-energy density of states,
that is intriguing in the context of the proposed deficit spending mechanism, and the local nature of the charges may be essential
for local tuning and control. Our work forms an element in resolving the story of
the nitrogenase enzyme, where many-electron wavefunction theory may now be used to access structures at the electronic level,
just as crystallography has advanced our understanding at the atomic level.

\section*{Author contributions}
Z.L. and G.K.C. designed the study and wrote the manuscript. Z.L. performed the calculations, supported by US NSF-CHE 1665333.
S.G. contributed to interfacing SP-MPS to SA-MPS used in the BLOCK code, supported by US NSF-SSI 1657286. Q.S. contributed to interfacing the PYSCF code with COSMO, supported by US NSF-SSI 1657286. All authors contributed to the discussion of the results.

\section*{Methods}
Starting from the P cluster structural models (see SM), we embedded all calculations
in a dielectric with $\epsilon=4.0$ to mimic the remaining protein environment.
For the 4 different model P clusters, we constructed the CAS using all Fe $3d$ and S $3p$ orbitals of the [\ceFeEight] core as well as the bonding orbitals with other ligands. This came to
 108 electrons in 71 orbitals (denoted as CAS(108e,71o)) for the synthetic cluster;
CAS(114e,73o) for \pnA, CAS(117e,75o) for \ppA, and CAS(120e,77o) for \poxA, respectively;
the size of the underlying many-electron Hilbert spaces range from $10^{31}$ to $10^{33}$ (Table S1).
Note that the different active spaces (arising from the change in ligation) for the different P-cluster states means that we cannot directly compare energies across different clusters; redox potentials will be the subject of future studies. The initial BS basins used
to initialize the DMRG calculations 
for each complex are summarized in Fig. 1b. In total, we examined 20 basins for the synthetic cluster ($S=0$), 35 for \pn ($S=0$),
70 for \pp ($S=1/2$ and $S=5/2$), and 152 for \pox ($S=3$ and $S=4$).
To understand how the energetic landscape arises, we computed local observables such as the charge
$N_{A}=\langle\hat{N}_A\rangle$ and local spin projection
$S_{A}^z=\langle\hat{S}_A^z\rangle$, as well as Fe-Fe spin-spin correlation functions $\langle\vec{S}_A\cdot\vec{S}_B\rangle$ (see SM).
From these, we deduced how the global states appear from recoupling local
orbital and spin degrees of freedom, and the relationships between the electronic states in \pnA, \ppA, and \poxA. The spin-projected and spin-adapted DMRG calculations were carried out
with the codes developed in Refs. \cite{spmps2017} and \cite{sharma_spin-adapted_2012}, respectively. Further
discussions of the  geometry, active spaces, accuracy, convergence, analysis, and other aspects, are in the SM.


%
\clearpage
\newpage

\clearpage
\newpage
\begin{figure}\centering
\begin{tabular}{cc}
\includegraphics[width=0.6\textwidth]{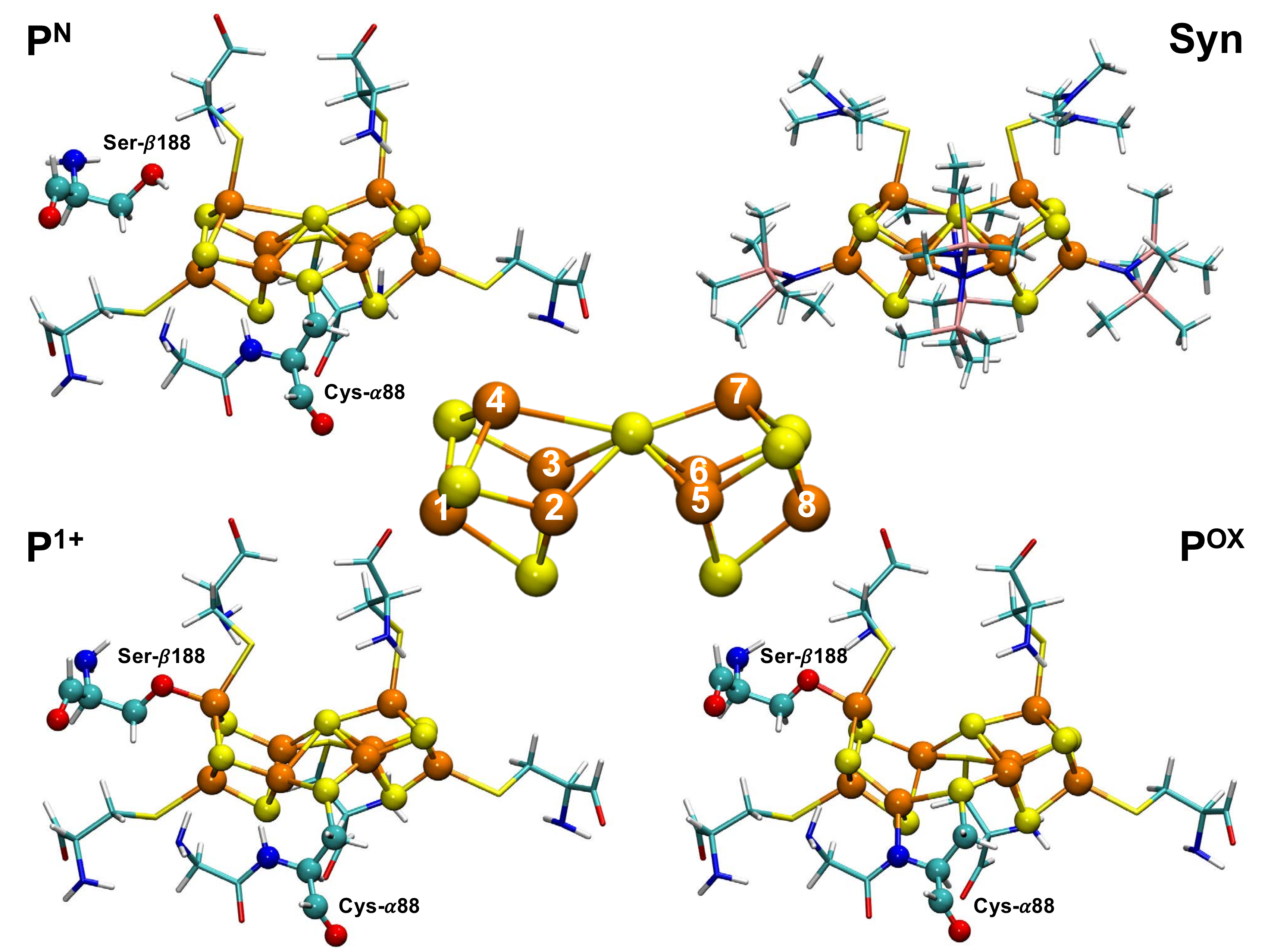} &
\raisebox{1.0cm}{\includegraphics[width=0.4\textwidth]{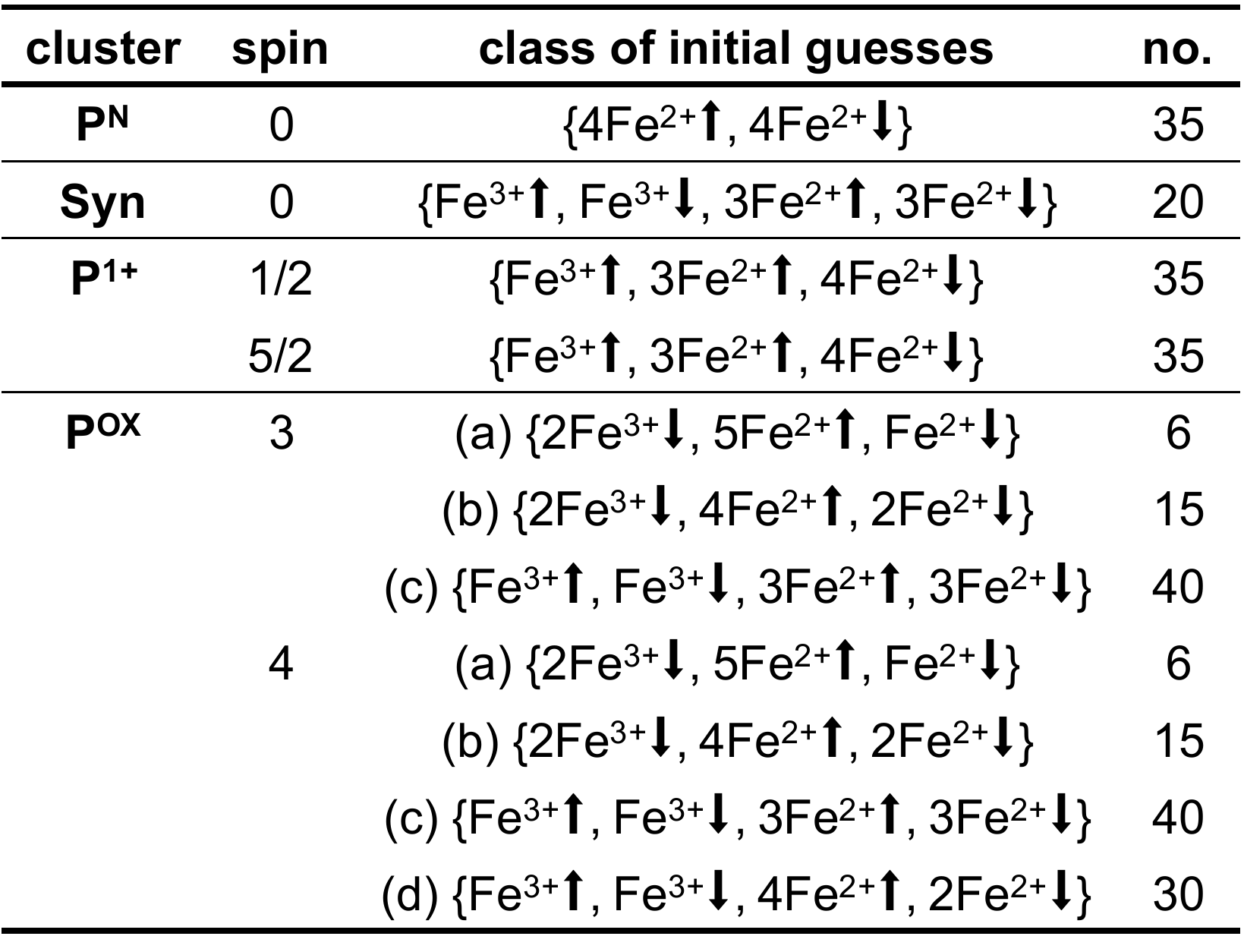}}\\
(a) & (b) \\
\multicolumn{2}{c}{\includegraphics[width=0.9\textwidth]{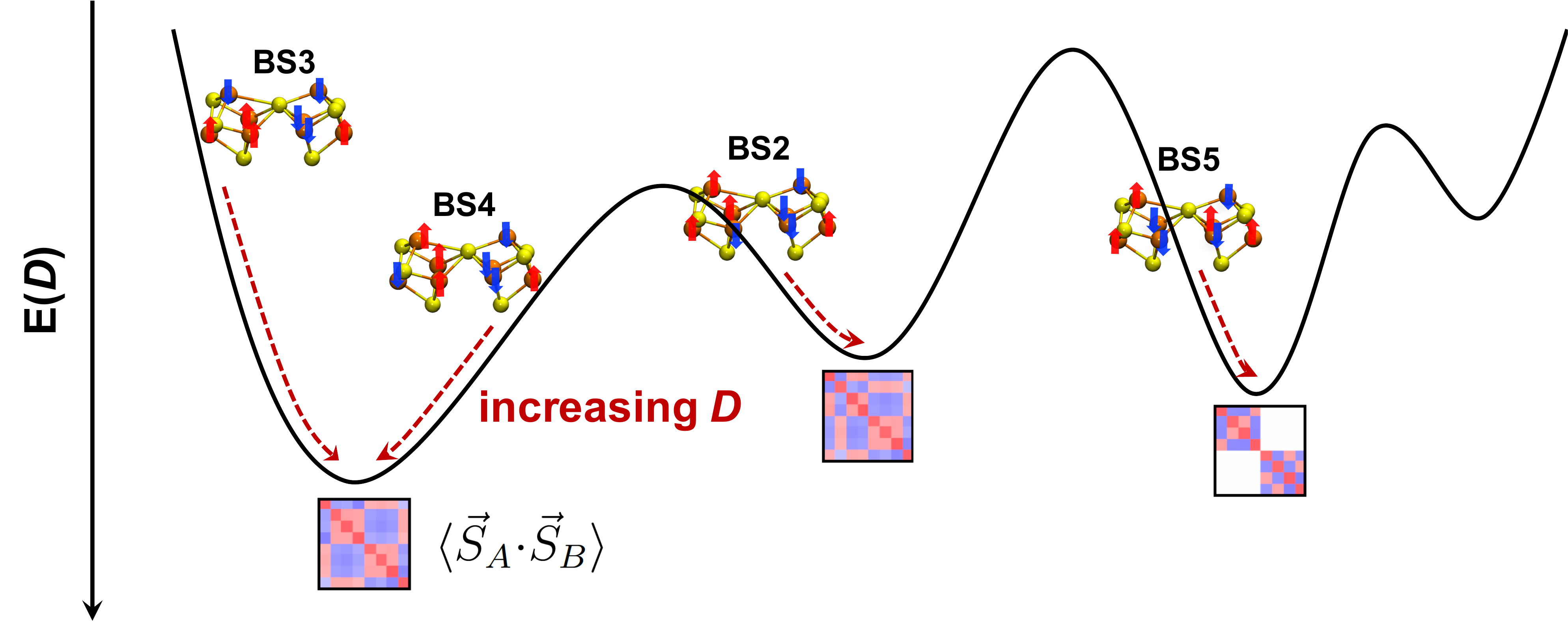}}\\
(c) \\
\end{tabular}
\caption{(a) [\ceFeEight] P-cluster models in the present study
(PDB ID: 3MIN for \pnA, 6CDK for \ppA, and 2MIN for \poxA,
and the synthetic analog of the \pn cluster in Ref. \cite{ohki2003synthesis}).
The labels in the central figure index the Fe atoms  in the discussion.
The redox-dependent structural rearrangement across \pnA, \ppA, and \pox involving Ser-$\beta$188 and Cys-$\alpha$88 is highlighted. Color legend: Fe, orange; S, yellow; C, cyan; O, red; N, blue; H, white; Si, pink. (b) Table of initial guesses (broken-symmetry product states) obtained by distributing different iron states
across the eight Fe atoms. (c) Schematic illustration of the change of energy
in the DMRG optimization process as a
function of bond dimension ($D$) starting from different broken-symmetry initial guesses.
The local minima in the parameter space represent (approximate) eigenstates of the
many-electron Schr{\"o}dinger equation within the active space.
They are characterized by the spin-spin correlation functions $\langle\vec{S}_A\cdot\vec{S}_B\rangle$ among eight irons (red: positive, blue: negative)
shown by the square graphs.}\label{fig:structures}
\end{figure}

\clearpage
\newpage
\begin{figure}\centering
\begin{tabular}{c}
\includegraphics[width=0.9\textwidth]{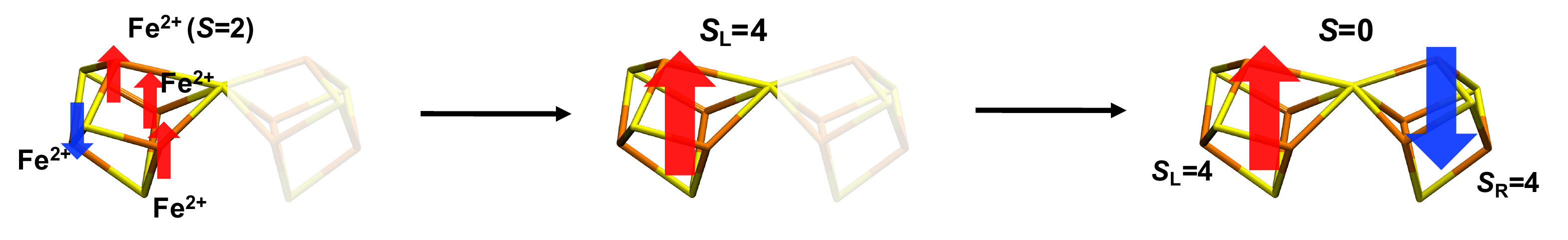} \\
(a) \\
\includegraphics[width=0.9\textwidth]{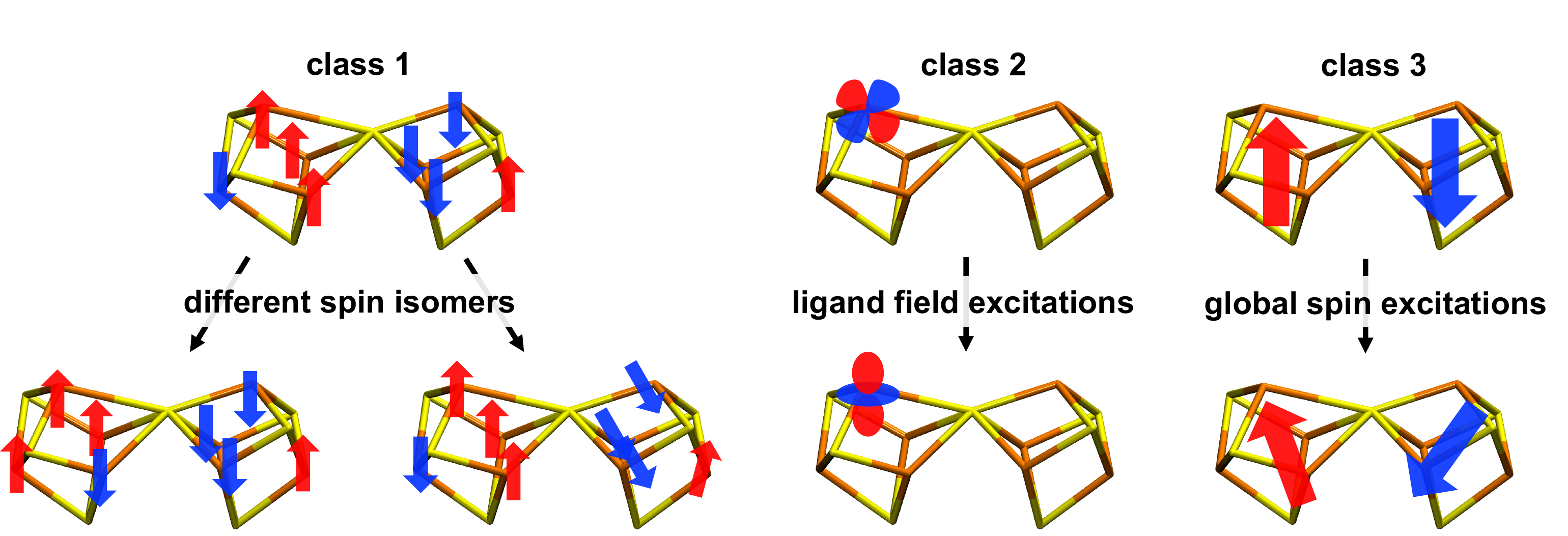} \\
(b) \\
\end{tabular}
\caption{(a) Illustration of the coupling of local states of Fe$^{2+}$ into
local cubane states ($S_L=4$) and
finally into the global $S=0$ ground state of \pnA.
(b) Schematic of typical excitations.
class A: spin isomers with the same spin but differ in the spin-couplings;
class B: ligand field (local orbital) excitations;
class C: global spin excitations that changes the global spins but
without altering the local cubane states.
}\label{fig:composition}
\end{figure}

\clearpage
\newpage
\begin{figure}\centering
\begin{tabular}{cc}
\includegraphics[width=0.5\textwidth]{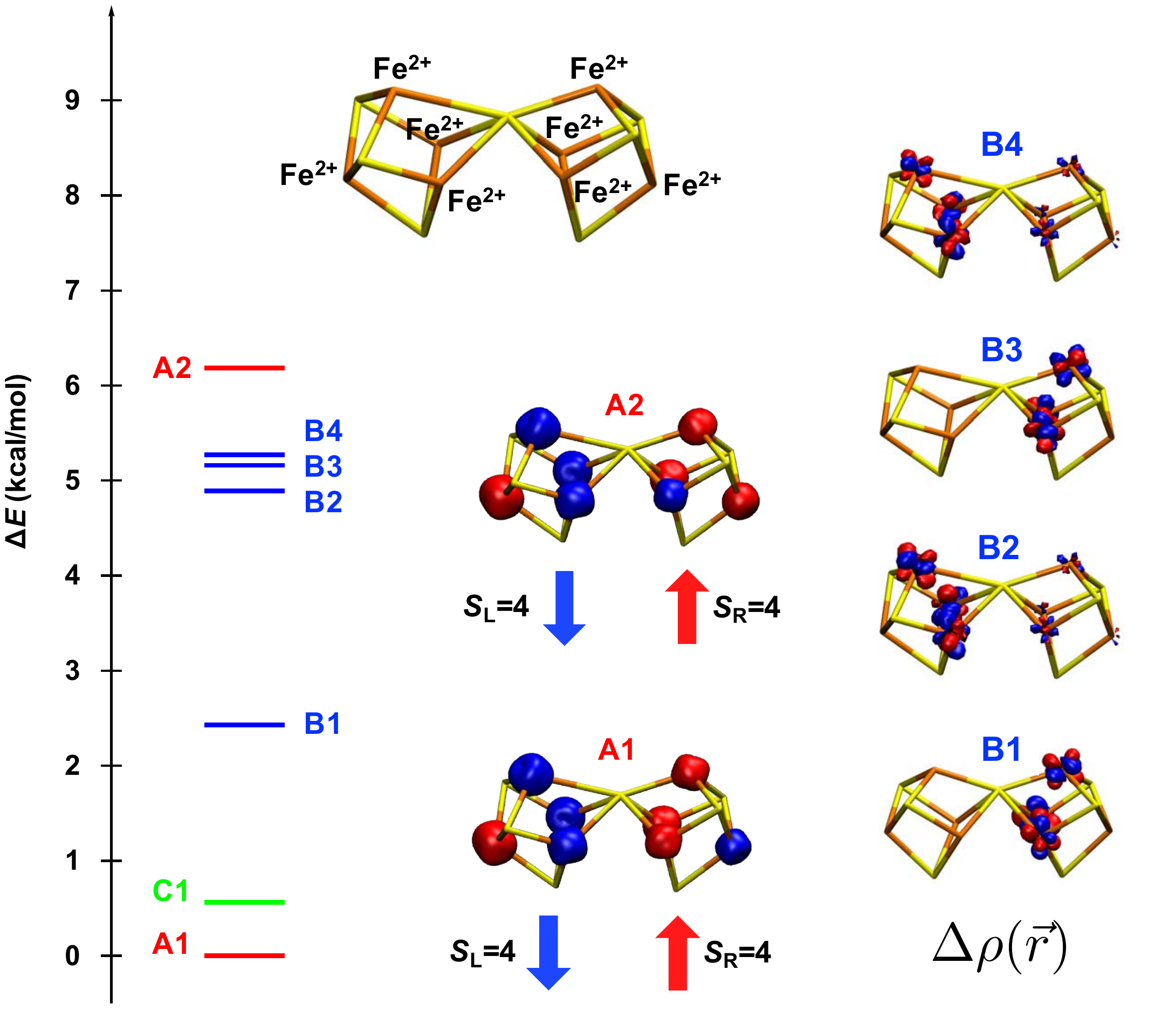} &
\includegraphics[width=0.5\textwidth]{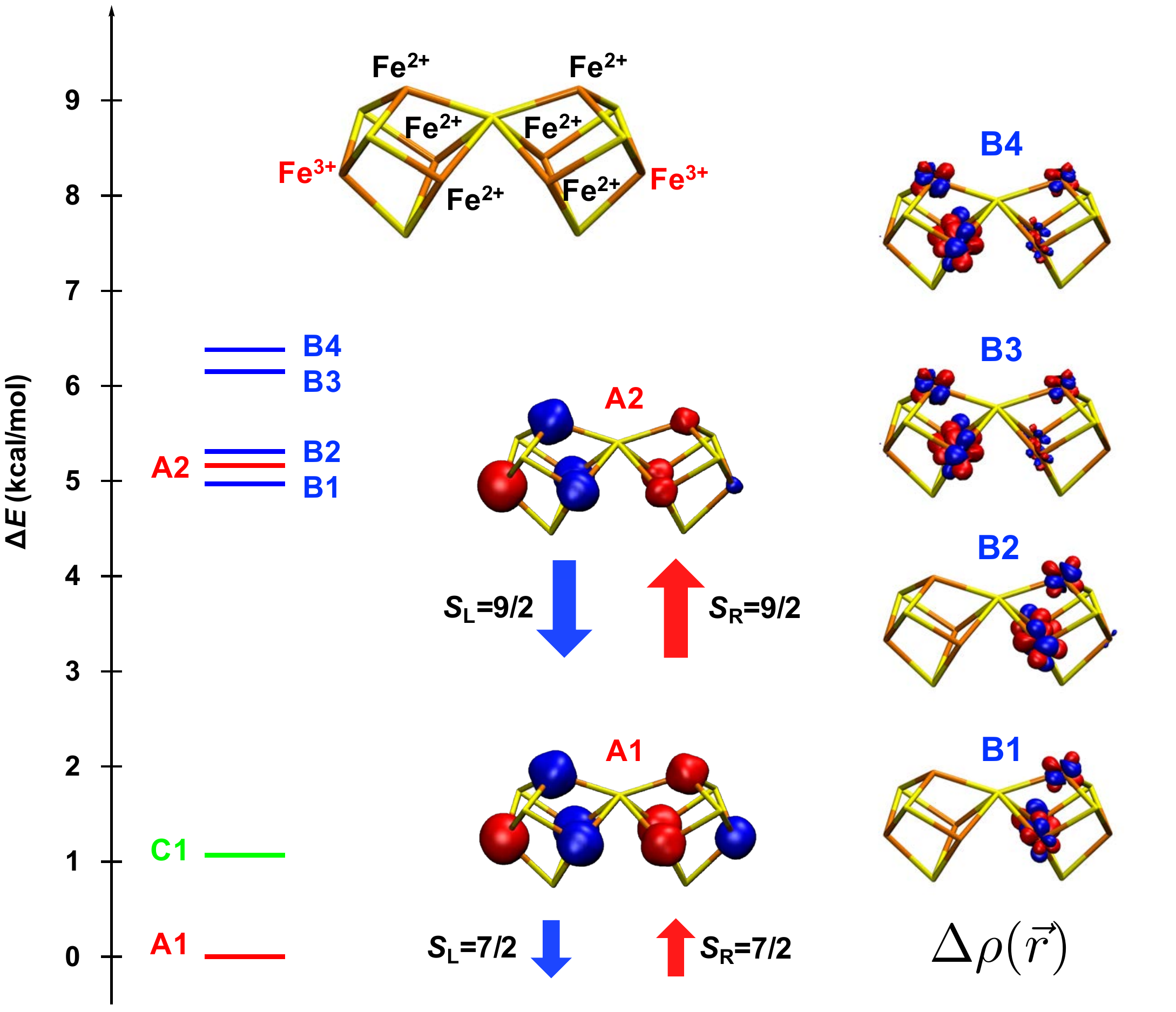} \\
(a) \pn ($S=0$)  & (b) Syn ($S=0$) \\
\end{tabular}
\caption{States of the (a)  \pn and (b)  synthetic cluster.
Left panel: extrapolated relative energies $\Delta E$ (kcal/mol) for spin isomers (red)
and estimated energies for the B1-B4 states via a state-averaged DMRG calculation
($D=3000$) starting from the converged ground state.
Inset: assigned formal oxidation states for the ground state based on local charge analysis.
Middle panel: spin-coupling patterns of states A1 and A2 revealed by
the spin-spin correlation density functions
$\sigma_{\mathrm{A}}(\vcr)=\langle \vec{S}_{A}\cdot\sum_{B}\vec{S}_{B}(\vcr)\rangle$
with A=Fe1 for the first iron of these two spin isomers.
A schematic representation of the spin-coupling scheme by a 'classical'
AFM coupling between two cubanes are shown below.
Right panel: density differences $\Delta\rho(\vcr)$ of B1-B4 states with respect to the ground state
density.}\label{fig:final_pnsyn}
\end{figure}

\clearpage
\newpage
\begin{figure}\centering
\begin{tabular}{cc}
\includegraphics[width=0.4\textwidth]{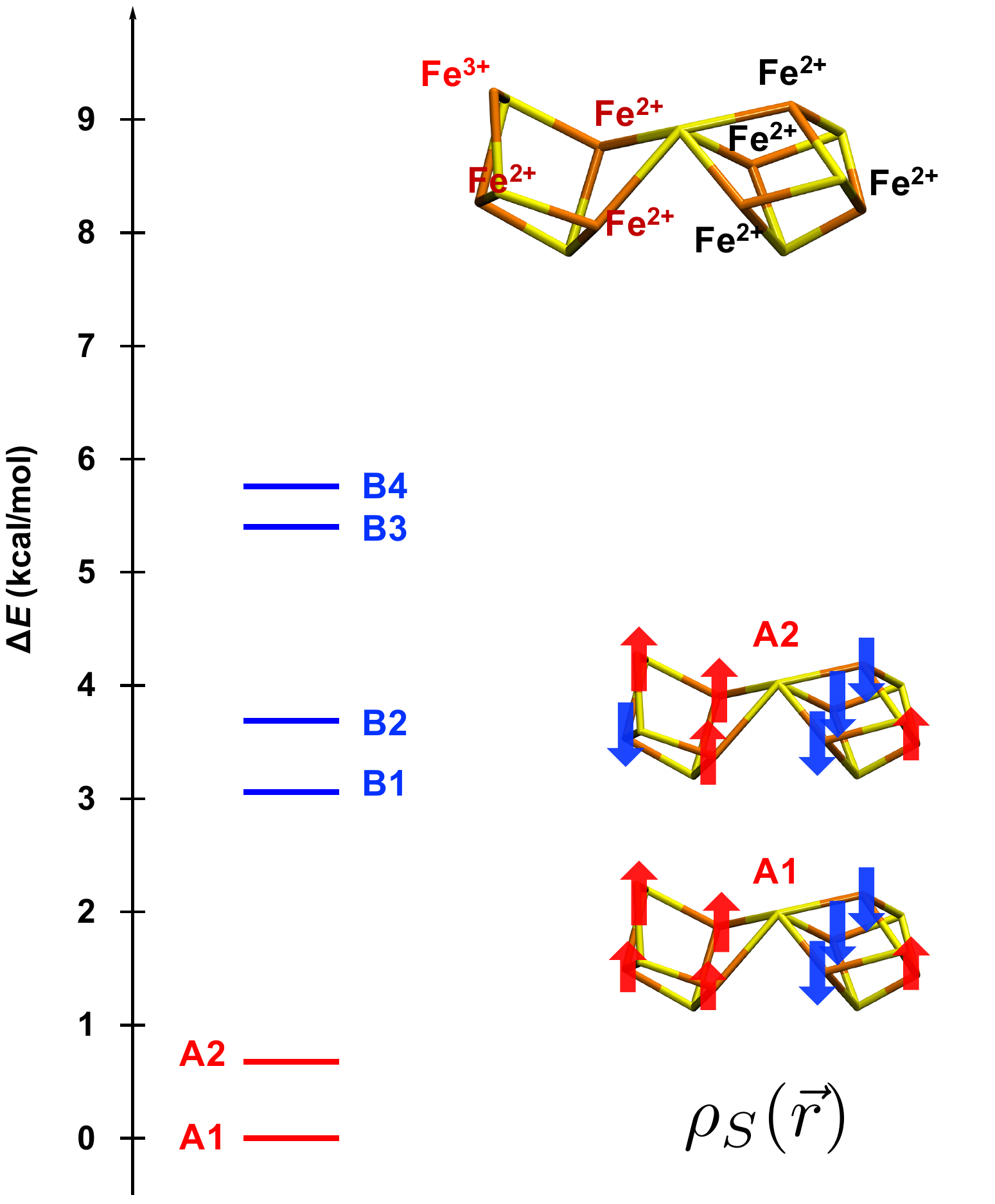} &
\includegraphics[width=0.4\textwidth]{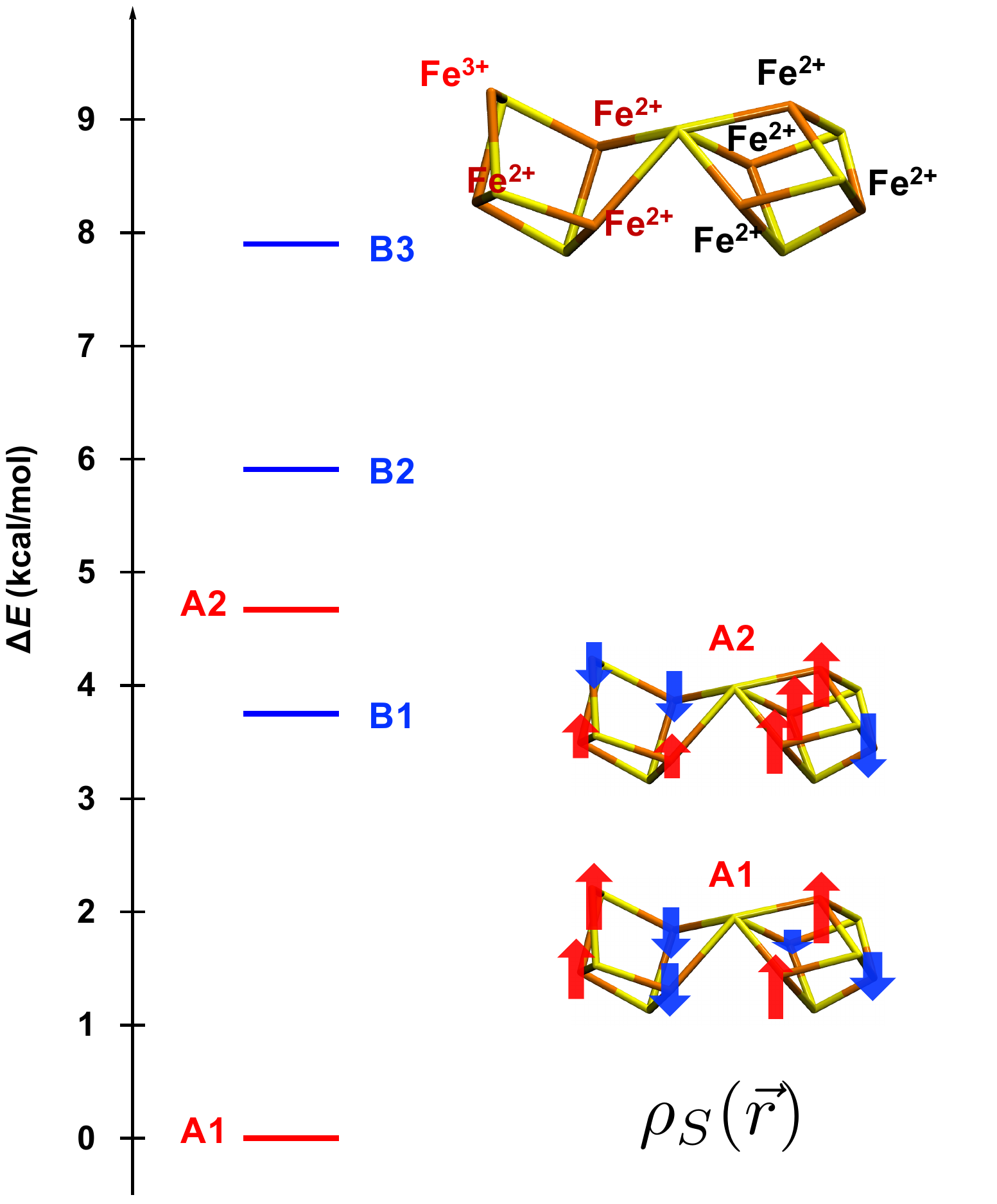} \\
(a) \pp ($S=1/2$) &
(b) \pp ($S=5/2$) \\
\includegraphics[width=0.4\textwidth]{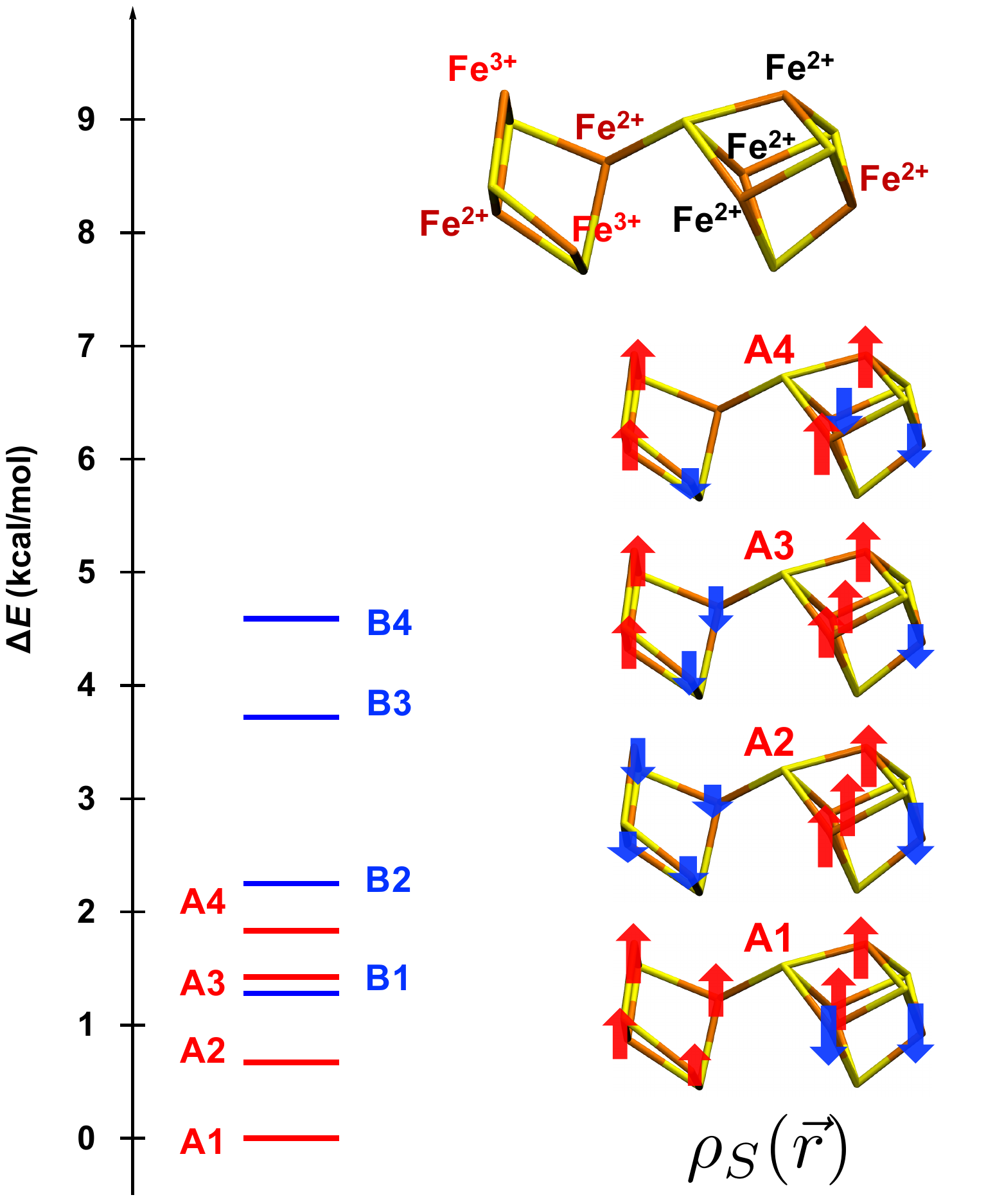} &
\includegraphics[width=0.4\textwidth]{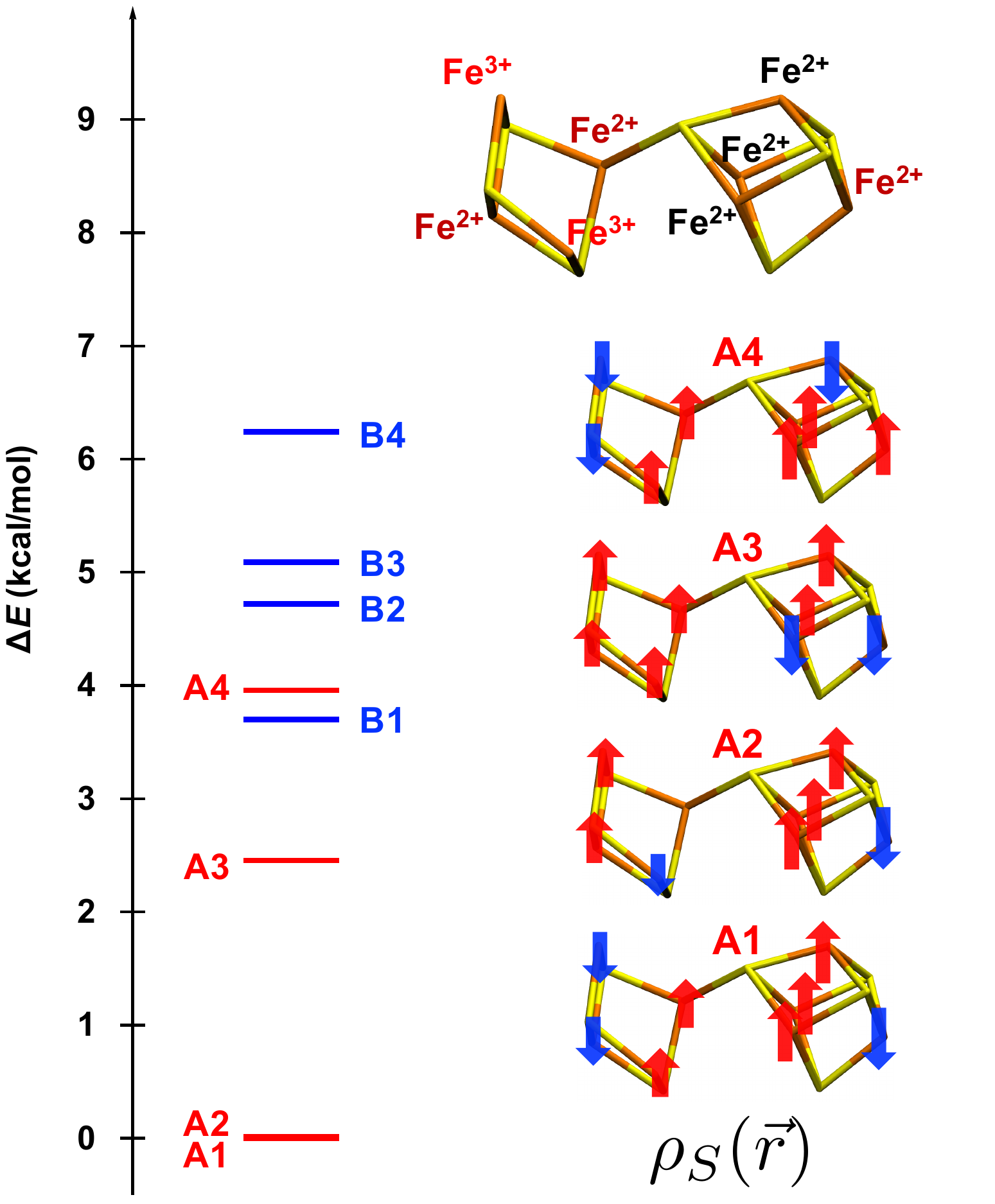} \\
(c) \pox ($S=3$) &
(d) \pox ($S=4$) \\
\end{tabular}
\caption{States of the oxidized P-clusters:
(a) \pp with $S=1/2$, (b) \pp with $S=5/2$,
(c) \pox with $S=3$, (d) \pox with $S=4$.
Left panel: see text in Fig. \ref{fig:final_pnsyn}.
The lowest $S=5/2$ state of \pp is lower than the lowest $S=1/2$ state by -2.7 kcal/mol
computed with extrapolated energies, while
the lowest $S=4$ state of \pox is lower than the lowest $S=3$ state by -1.4 kcal/mol
computed with extrapolated energies.
Inset: assigned formal oxidation states for the ground state based on local charge analysis.
Fe$^{2+}$ symbols in dark red indicate that the irons have ferric character due to charge delocalization as compared to other ferrous irons in black.
Right panel: schematic representation of the spin-density $\rho_S(\vcr)$ of the lowest spin-isomers.
}\label{fig:final_pox}
\end{figure}

\clearpage
\newpage
\begin{figure}\centering
\includegraphics[width=1.0\textwidth]{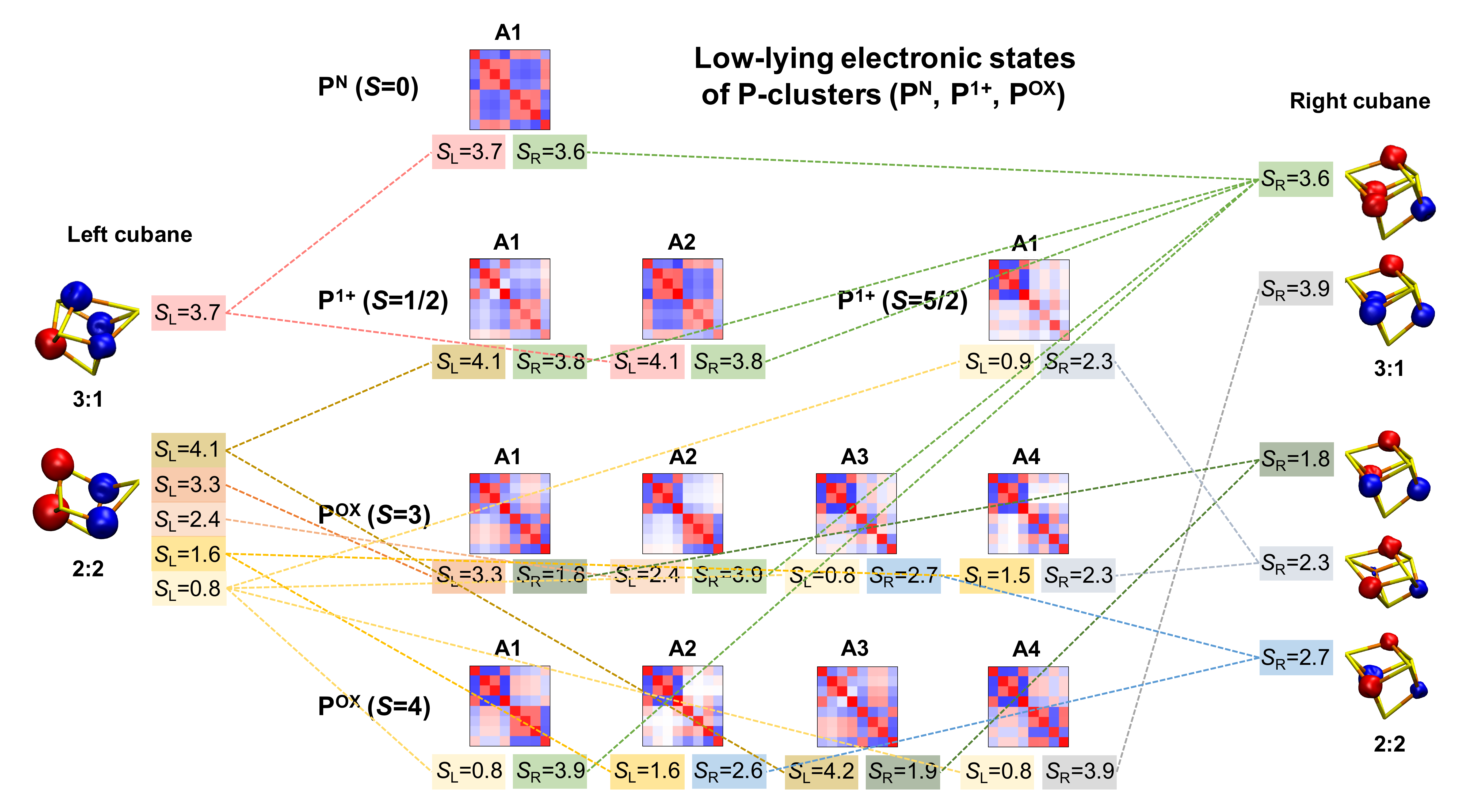}
\caption{Relationships between spin-isomers in low-lying electronic states of the P-clusters,
which unravel the evolution of local (left or right) cubane states upon oxidations.
The cubane states sharing the same color are basically the same spin state.
Square graph: Spin-spin correlation functions $\langle\vec{S}_A\cdot\vec{S}_B\rangle$ among eight irons (red: positive, blue: negative). The effective spins of the left cubane $S_L$ and the right cubane $S_R$ are computed for the irons on the left and right cubanes, respectively.}\label{fig:map}
\end{figure}

\end{document}